\documentclass[showpacs,twocolumn,aps,prb]{revtex4}
\usepackage{amsmath}
\usepackage{amsfonts}
\usepackage{amssymb}
\usepackage{amsthm}
\usepackage{graphicx}

\begin{document}

\title{Proximity and anomalous field-effect characteristics
in double-wall carbon nanotubes}

\author{Jie Lu}
\affiliation{Physics Department, The Hong Kong University of
Science and Technology, Clear Water Bay, Hong Kong SAR, China}
\author{Sun Yin}
\affiliation{Physics Department, The Hong Kong University of
Science and Technology, Clear Water Bay, Hong Kong SAR, China}
\author{L. M. Peng}
\affiliation{Key Laboratory for the Physics and Chemistry of
Nanodevices and Department of Electronics, Peking University,
Beijing 100871, China}
\author{Z. Z. Sun}
\affiliation{Physics Department, The Hong Kong University of
Science and Technology, Clear Water Bay, Hong Kong SAR, China}
\author{X. R. Wang}
\email[To whom correspondence should be addressed. Electronic
address: ]{phxwan@ust.hk}
\affiliation{Physics Department, The Hong Kong University of
Science and Technology, Clear Water Bay, Hong Kong SAR, China}
\date{\today}

\begin{abstract}
Proximity effect on field-effect characteristic (FEC)
in double-wall carbon nanotubes (DWCNTs) is investigated.
In a semiconductor-metal (S-M) DWCNT, the penetration of
electron wavefunctions in the metallic shell to the
semiconducting shell turns the original semiconducting tube
into a metal with a non-zero local density of states at the
Fermi level. By using a two-band tight-binding model on a
ladder of two legs, it is demonstrated that anomalous FEC
observed in so-called S-M type DWCNTs can be fully understood
by the proximity effect of metallic phases.
\end{abstract}


\maketitle
{\it Introduction--} Carbon nanotubes (CNTs) have attracted much
attention in recent years because of their novel properties\cite
{dresselhaus} and potential applications in devices\cite{chico}
and wiring\cite{sander1} in nano and molecule electronics.
The recent advance in technology allows the fabrication of both
single wall carbon nanotubes (SWCNTs) and multi-wall carbon
nanotubes (MWCNTs) with controllable diameters. Depending on its
chirality\cite{dresselhaus}, a SWCNT can be metallic or
semiconducting. Various issues\cite{lambert,mayou}
have been examined, including electronic transport\cite
{sander2,avouris,dai,shimada,shidong} of different types of CNTs.

Very recently, field-effect transistors have been fabricated
out of double-wall carbon nanotubes (DWCNTs) which are
the simplest MWCNTs. Field-effect characteristic (FEC), which
is about how source-drain current $I_{sd}$ depends on gate
voltage $V_G$, has been measured\cite{Peng}. Each DWCNT devices
can experimentally be classified into one of three groups,
corresponding to semiconductor-semiconductor (S-S), metal-metal
(M-M) or metal-semiconductor (M-S), and semiconductor-metal
(S-M) combinations of two shells of DWCNTs (the first symbol
is for the outer shell and the second for the inner one).
The S-S and M-M (or M-S) DWCNTs exhibit similar FEC as those of
semiconducting and metallic SWCNTs\cite{sander2}, respectively.
However, S-M DWCNT devices show FEC with distinct features
from both of its metal and semiconductor counterparts.
1) In the negative $V_G$ region, on-off current ratio
($I_{on}/I_{off}$) can be as large as $10^1$ or even $10^2$.
This is in contrast with no obvious switching characteristic for
metallic tubes or order of $10^5$ on-off ratio for semiconducting
ones. 2) The gate-voltage dependence of source-drain current is
not exponential in transition region from on to off states as
one will expect from the orthodox semiconductor physics theory.
3) The transition region is much wider than that of usual
semiconductors. 4) For positive $V_G$, it seems that
free charges in the metallic tube may screen the outer shell
from the gate effect, harmful to field-effect transistors.

In this paper, we would like to attribute the anomalous FEC
observed in the so-called S-M DWCNTs to the {\it proximity
effect} of metallic phase, a general property of waves.
A qualitative explanation of the anomalous FEC is presented
first. The penetration of electron wavefunctions in the inner
metallic tube into the outer semiconducting shell due to the
inter-shell coupling generates local electron density of states
(DOS) in the semiconducting tube in its original forbidden gap.
Thus, new conducting channels are created in the gap,
and the semiconducting tube becomes weakly conducting.
The picture is verified by modeling a DWCNT by a two-band
tight-binding Hamiltonian on a ladder of two legs.
The results capture the essential features of the observed FEC.

\begin{figure}[htbp]\centering
\scalebox{0.26}[0.26] {\includegraphics[angle=0]{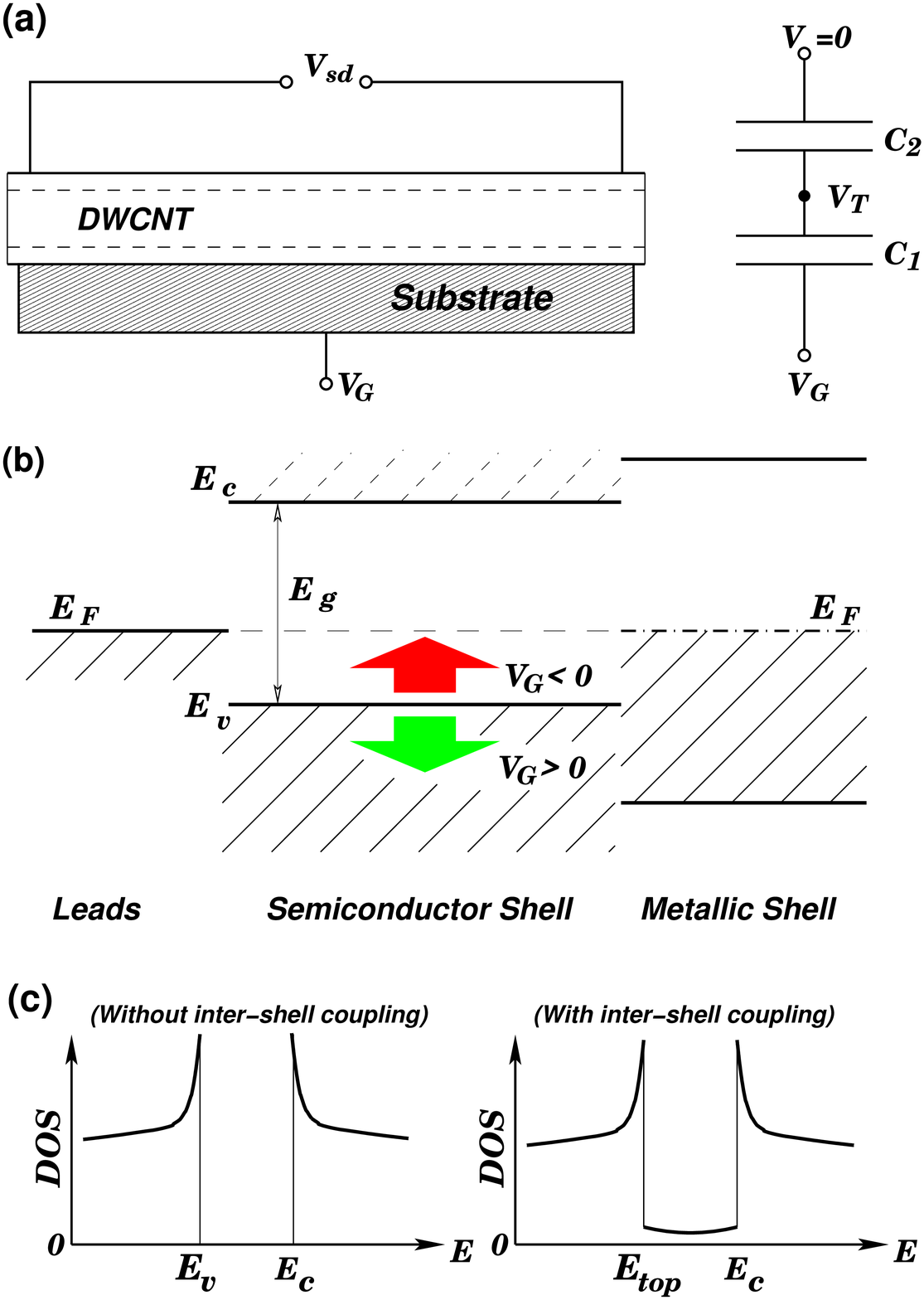}}
\caption{(a) Schematic experimental set-up of measuring FEC of a
DWCNT. $V_{sd}$ and $V_G$ are source-drain bias and gate voltage,
respectively. $C_1$ and $C_2$ are the capacitance between the gate
and the DWCNT and that of the DWCNT itself. $V_T$ is the potential
on the tube. (b) Schematic energy bands of the outer semiconducting
shell (middle part), and the conduction band of the inner metallic
shell (right part). $E_F$ is the Fermi level of the metallic leads.
$E_c$ ($E_v$) is the bottom (top) of the conduction (valence) band
and $E_g$ is the energy gap. The big arrows indicate the moving
directions of $E_v$ under $V_G$. (c) The DOS in the outer shell
without and with an inter-shell coupling which leads to a non-zero
DOS in the gap region.}\label{fig1}
\end{figure}
{\it Picture--} Consider a S-M DWCNT under a gate voltage $V_G$
schematically shown in Fig.\ref{fig1}a, a source-drain bias
$V_{sd}$ is applied between the two ends of the outer DWCNT shell.
The inner and outer tubes are assumed to be metallic and
semiconducting, respectively. Their energy band structures are
like those of the right and the middle parts of Fig.\ref{fig1}b.
Without an inter-shell coupling, the system consists of two
independent tubes with their own energy bands. In the absence of
the gate voltage, the DOS in the outer shell at the Fermi level
is zero. Thus, electrons (holes) can move from the source to the
drain through either the inner metallic shell by quantum tunneling or
the outer-shell conduction (valence) band by thermal activation.
Both process contribute to the leakage current in field-effect
transistors, and are small at low temperature with $kT\ll E_g$.
Applying a negative gate voltage, the top of the valence band
will be pushed upward (shown by the big arrows in  Fig.~
\ref{fig1}b) so that hole conduction will increase exponentially
with the gate voltage, leading to the FEC of a usual semiconductor
device.

In reality, there is always an inter-shell coupling in a DWCNT.
Electron states in two shells are hybridized with each other such
that the local DOS of the two shells will be modified (the total
DOS may not change much). For example, the local DOS of the outer
semiconducting shell in its original energy gap becomes non-zero
because inner-shell states in the energy range have certain
probability to appear on the outer shell. The penetration
of electron wavefunctions from the metallic shell into the
semiconducting shell is called the {\it proximity effect}. For
comparison, the DOS of the outer shell with and without inter-shell
coupling are illustratively plotted in Fig.\ref{fig1}c.
Its value in the original gap region is sensitive to the energy
barrier $E-E_v$ for the electrons in the inner shell, where $E$
and $E_v$ are the electron energy and top of the valence band,
respectively. This barrier can be controlled by the gate voltage
$V_G$ because $E_v$ is pushed upward under a negative $V_G$,
resulting in the decrease in the barrier and increase in the
local DOS of the outer shell.

The appearance of DOS in the outer shell in the
energy gap creates new electron conducting channels.
At a small $V_G$ and under a small $V_{sd}$, all current can
be carried by the local states in the outer shell, and there is
no need to use the inner shell although the DOS there near the
Fermi level is much greater than that in the outer one. This
feature has been observed in MWCNT\cite{frank,abellan,lambert}.
The proximity effect, which becomes stronger and stronger as the
gate voltage is more and more negative, turns the semiconducting
outer shell into a weak metal. The gate-voltage dependence
of this metallicity leads to the anomalous FEC in S-M DWCNTs.
Since the origin of the anomaly is from the emergence of DOS in
the outer shell in its gap region, the proximity effect is
important only when the Fermi level lies in the gap, i.e. $E_v<
E_F<E_c$. This explains the relative wide transition region, as
well as the non-exponential dependence of $I_{sd}$ on $V_{G}$.

\begin{figure}[htbp]\centering
\scalebox{0.32}[0.32]{\includegraphics[angle=0]{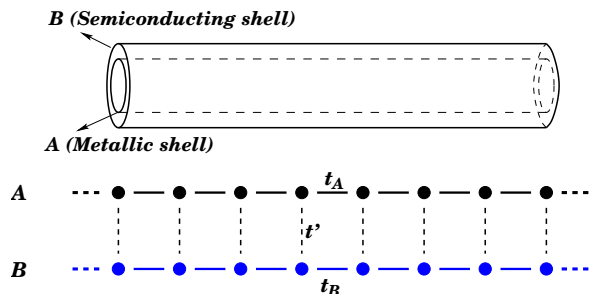}}
\caption{A S-M DWCNT (top) is modeled by a two-leg ladder (bottom)
on which a tight-binding model is defined. $t_A (t_B)$ is the
hopping coefficient in leg A (B), and $t'$ is the inter-chain
hopping coefficient.}\label{fig2}
\end{figure}

{\it Model--} To put the above picture on a firm quantitative
analysis, a S-M DWCNT is modeled by a two-leg ladder as shown in
Fig.\ref{fig2}. One leg, say leg $A$, is used to model the inner
metallic shell, and the other, leg $B$, is for the outer semiconducting
shell. If the source-drain bias is not very large, then there is
only one band from each shell relevant to electronic transport.
Following the experimental situation\cite{Peng}, it is assumed that
the semiconducting shell is originally $p$-doped so that its Fermi
level is near the top of the valence band. However, it is trivial
to generalize the following discussion to an $n$-doped tube where
the conduction band will be considered and the band bottom is
slightly above the Fermi level. A tight-binding model is defined
on the ladder with $t_A$ and $t_B$ (order of the shell bandwidths)
being the hopping coefficients on leg $A$ and $B$, respectively.
The inter-shell coupling is described by inter-chain hopping
coefficient $t'$. The Hamiltonian is
\begin{equation}\label{hamilton}
H=\sum_{\alpha,i}(\frac{1}{2}\varepsilon_{\alpha}c_{\alpha,i}^{\dagger}
c_{\alpha,i}+t_{\alpha}c_{\alpha ,i}^{\dagger}c_{\alpha,i+1})
+\sum_{i}t'c_{A,i}^{\dagger}c_{B,i}+h.c.,
\end{equation}
where $\alpha=A,B$ labels the legs, $\varepsilon_{\alpha}$ are
the on-site energies. $c_{\alpha,i}^{\dagger}$ and $c_{\alpha,i}$
are the creation and annihilation operators at site $i$ of leg
$\alpha$, respectively. The Hamiltonian has two branches of
eigen-energies,
\begin{eqnarray}\label{new2bands}
E_{\pm,k}& = & \frac{1}{2}(\varepsilon_A+\varepsilon_B)+(t_{A}+
t_{B})  \cos k\\
& & \pm \frac{1}{2}\sqrt{[(\varepsilon_A-\varepsilon_B
)+2(t_{A}-t_{B}) \cos k ]^{2}+4|t'|^{2}}. \nonumber
\end{eqnarray}
Without inter-chain coupling, i.e. $t'=0$, the two branches
correspond to two independent bands (width $4t_A$ and $4t_B$)
centered at $\varepsilon_A$ and $\varepsilon_B$, respectively.
In general, $\varepsilon_A$ and $\varepsilon_B$ depend on the
gate voltage. However, $\varepsilon_A$ is less sensitive to
$V_G$ because DOS in the inner shell at the Fermi level is
large, and any substantial change of $\varepsilon _A$ results
in a great accumulation of charge in the shell which will
generate a large electric potential. This is also often called
Fermi level pining (at large DOS). Therefore, the gate voltage
applies mainly on the outer shell, and it is reasonable to
assume a constant $\varepsilon_A$ and $\varepsilon_B=U_0-e\beta
V_G$, where $U_0$ is the initial on-site energy and $\beta
=C_1/(C_1+C_2)$ relates to the mutual capacitance $C_1$ between
the substrate and the DWCNT and the self-capacitance $C_2$
of DWCNT as shown in Fig. \ref{fig1}a. For a given DWCNT (fixed
$C_2$), larger $C_1$ means larger $\beta$. $\beta$ increases as
the substrate thickness decreases because $C_1$ becomes bigger.
This explains why $I_{sd}$ under the same $V_{sd}$ is bigger
with a thinner $SiO_2$ layer\cite{Peng} since a thin substrate
is more effective in applying a gate voltage. Without losing
generality, the Fermi level of the source lead is set to zero.
Model parameters are chosen so that $E=0$ is inside the
energy band of leg-A but outside that of leg-B at $V_G=0$.
\par

As it was mentioned early, the important quantity is the DOS
in leg $B$ which can be evaluated by
\begin{equation}\label{dos-expression}
g_B(E)=\frac{1}{2\pi}\sum_{\lambda=\pm,k} \left|\frac{\partial
E_{\lambda,k}}{\partial k}\right|^{-1}P_{\lambda,k,B}
\delta(E_{\lambda,k}-E),
\end{equation}
where $P_{\lambda,k,B}\equiv \sum_{i_B}
\left|\psi_{\lambda,k}(i_B)\right|^{2}$ is the probability of
finding an electron in leg $B$ when it is in eigenstate
$\psi_{\lambda,k}(x)$.

\begin{figure}[htbp]\centering
\scalebox{0.7}[0.7]{\includegraphics[angle=0]{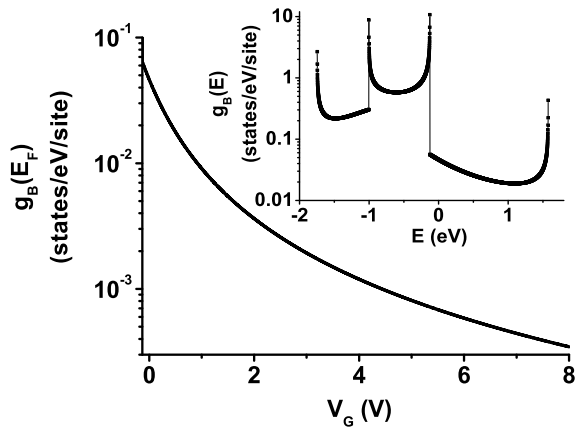}}
\caption{The logarithm of $g_B(E_F)$ vs. gate voltage $V_G$ when
$\varepsilon_B=U_0-eV_G$ is used.
Inset: The logarithm of $g_B(E)$ vs. $E$ at $V_G=0$.}\label{fig3}
\end{figure}

{\it Results and discussions--} To have a better picture about
how large the proximity effect is and to make a quantitative
comparison with experimental data, the model parameters will be
set as follows: $t_A=0.75eV$; $t_B=0.3eV$; $\varepsilon_A
=0eV$; $U_0=-.65eV$; $t'=0.35eV$\cite{mayou,michenaud},
corresponding to a typical subband width of $1eV \sim 4eV$ of
CNTs\cite{mayou,michenaud}, and the top of the valence band of
the outer shell being about $0.05eV$ below $E_F$ at $V_G=0$.
Thus, $E_F=0$ is inside the band of leg-A. The inset of
Fig.{\ref{fig3}} is $g_B(E)$ as a function of energy $E$
(in $eV$) at $V_G=0$. The DOS in the energy range of $-1.75eV<
E<-0.13eV$ is the sum of the contribution from `-' branch of
\eqref{new2bands}, a 1-D tight-binding model like, plus the
proximity effect contribution from `+'-branch of
\eqref{new2bands} in the range of $-1eV\leq E\leq -0.13eV$.
Its typical value is of order of 1/($eV\cdot$site).
The non-zero DOS outside of this energy range is completely
from the proximity effect of `+'-branch of \eqref{new2bands}.
The spikes are the van Hove singularities of the 1-D
tight-binding model, a model-dependent feature.

Fig.{\ref{fig3}} is the local DOS $g_B(E_F)$ in leg-B at the
Fermi level as a function of gate voltage $V_G$ (in volts $V$).
The model parameters are the same as those in the inset of Fig.
{\ref{fig3}}, and $\varepsilon_B=U_0-eV_G$ ($\beta=1$) is
used for simplicity. For large $V_G$, the Fermi level is deep
inside the original energy gap, and the value of $g_B(E_F)$ is
small, order of $10^{-3}$($eV\cdot$site)$^{-1}$. The Fermi level
approaches the top of valence band when $V_G$ becomes negative.
Near $V_G=-0.13V$, the DOS is order of $10^{-1}$($eV\cdot$site)
$^{-1}$, a order of $10^2$ change in electronic conduction as
$V_G$ varies.

\begin{figure}[htbp]\centering
\scalebox{0.70}[0.70]{
\includegraphics[angle=0]{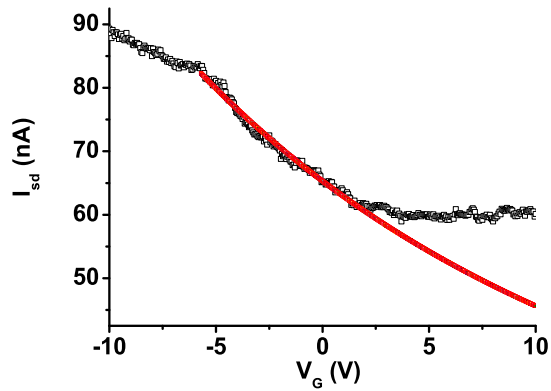}}
\caption{Fitting between the analytical form (solid line) of the
local DOS in the outer (semiconducting) shell at Fermi level
and the experimental data (Ref. 11) at $292K$. The saturated
$I_{sd}$ (around $60nA$) is interpreted as the leakage current.
Good fit indicates applicability of the simple model.}\label{fig4}
\end{figure}

Under a small source-drain bias $V_{sd}$, the current $I_{sd}$
is proportional to the local DOS on the outer shell at the Fermi
level. Fig. \ref{fig4} is the fitting curve of experimental FEC
at temperature ($292K$)\cite{Peng} by $g_B(E_F)$ for $V_G\ge -5V$,
where proximity effect dominates the electron transport process.
Here, it is interpreted that the top of the valence band touches
the Fermi level at $V_G=-5V$ in the experiment\cite{Peng}.
Thus both of the states from the proximity effect and from the
orginal valence band particpated in the transport.
This is why a reflexion point occurs in the experimental $I_{sd}$
curve at $V_G=-5V$, and the present theory should only compare
with the data for $V_G\ge -5V$. In the figure, $\varepsilon_A -
\varepsilon_B=-U_0+\beta e V_G$ and $I_{sd}=\eta g_B$ are used.
The fitting parameters are $t_A=0.95eV$; $t_B=0.44eV$; $t'=0.42eV
$; $\varepsilon_A=0.11$; $U_0=-1.05eV$; $\beta=0.025$; and $\eta
=3.1\mu$A$\cdot eV$. These values are reasonable for CNTs, and it
is surprising that these values can be obtained by fitting the
experiments to such a simple model. The fitting is not bad for
$I_{sd}\ge 60$nA. The deviation for $I_{sd}<60$nA can be
attributed to the leakage current or the gate-voltage dependence
of $\beta$ due to screening effect. It is not difficult to
understand the deviation because the proximity effect does not
take into account the leakage current and the screening effect.
The leakage current is very sensitive to the temperature as
what have been observed in experiment\cite{Peng}.

$I_{on}/I_{off}$ is sensitive to the temperature.
At zero temperature, $I_{off}$ comes completely from the
proximity effect through the appearance of the local DOS in
the outer shell in the gap region. Increase the temperature,
an extra current from electrons (holes) in the conduction
(valence) band of the outer shell is added to $I_{off}$.
This extra current could be comparable or even bigger than
that of the proximity effect at high enough temperature.
This extra leakage current will also add to $I_{on}$, but
it will be less important because of relative large current
of $I_{sd}$ in the on-state. In general, this ratio
decreases with the increase of the temperature, exactly as
what was found in experiments\cite{Peng}.

It should be emphasized that the anomalous FEC observed in
S-M DWCNTs is not due to change of holes (or electrons in
the conduction band of semiconductors) as that of usual
semiconductor field-effect transistors under a gate voltage.
Rather, it is due to change of DOS originated from the
proximity effect at the Fermi level. Although the detail
atomic and energy-band structures of DWCNT are neglected,
the present model reveals the essential physics of the
anomalous FEC of S-M DWCNTs. It implies that the anomaly is a
common feature of all S-M layered structures, not necessary
for CNTs only. This is because the proximity effect is a
general property of wave nature of particles, independent
of system details. The proximity effect is not new.
It is an important effect in superconductivity that has
received intensive investigation. The Josephson tunneling is
one of its manifestations. The difference of the proximity
effect in superconductors and in DWCNTs is that the particles
are Cooper pairs in the former and electrons in the latter.

Electron-electron (e-e) interactions were neglected in this study.
In general, e-e interactions are important in low temperature
physics and for the narrow bands. For CNTs, the band width is
quite large (order of $1\sim 4eV$) so that the net e-e
interaction may be reduced through screening. So far, most
physics quantities measured in experiments have not shown the
importance of the e-e interactions. Thus, it is safe to neglect
the interaction for the anomalous FEC which was measured at $292K$.
However, the interaction may be very important for very
low temperature physics and very small wires where the Luttinger
liquid (non Fermi liquid) behavior will dominate eventually.


{\it Conclusion--} The anomalous FECs in so-called S-M DWCNTs
is explained in terms of one-electron energy band structure.
The essential physics is the proximity effect of electrons:
The proximity effect creates local DOS in the gap region,
resulting in new conduction channels in the semiconducting outer
tube. The proximity effect is stronger and stronger when the
Fermi level approaches the top (bottom) of the valence (conduction)
band of the outer p-doped (n-doped) semiconducting tube.
As a result, the semiconducting outer tube becomes more and more
metallic.

{\it{Acknowledgments}--}This work is supported by UGC, Hong Kong,
through RGC CERG grants (No. HKUST6067/02P; 603904).
LMP is supported by NSF of China (Grant No. 10434010).

\end{document}